\documentclass[12pt]{iopart}
\usepackage{tensor}
\usepackage{iopams}
\usepackage{float}
\usepackage{tikz}
\usetikzlibrary{backgrounds}
\usepackage{subcaption}
\begin{document}
\title[Pancakes as opposed to Swiss Cheese]{Pancakes as opposed to Swiss Cheese}

\author{S N\'ajera$^1$, R A Sussman$^1$}

\address{$^1$ Instituto de Ciencias Nucleares, Universidad Nacional Aut\'onoma de M\'exico
(ICN--UNAM),\\ A. P. 70 –- 543, 04510 M\'exico D. F., M\'exico.}
\ead{sebastian.najera@correo.nucleares.unam.mx}

\begin{abstract}
We examine a novel class of toy models of cosmological inhomogeneities by smoothly matching along a suitable hypersurface an arbitrary number of sections of ``quasi flat'' inhomogeous and anisotropic Szekeres-II models to sections of any spatially flat cosmology that can be described by the Robertson--Waker metric (including de Sitter, anti de Sitter and Minkowski spacetimes). The resulting ``pancake'' models are quasi--flat analogues to the well known spherical ``Swiss-cheese'' models found in the literature. Since Szekeres-II models can be, in general, compatible with a wide range of sources (dissipative fluids, mixtures of non--comoving fluids, mixtures of fluids with scalar or magnetic fields or gravitational waves), the pancake configurations we present allow for a description of a wide collection of localized sources embedded in a Robertson--Waker geometry. We provide various simple examples of arbitrary numbers of Szekeres-II regions (whose sources are comoving dust and energy flux interpreted as a field of peculiar velocities) matched with Einstein de Sitter, $\Lambda$CDM and de Sitter backgrounds. We also prove that the Szekeres--II regions can be rigorously regarded as ``exact'' covariant perturbations on a background defined by the matching discussed above. We believe that these models can be useful to test ideas on averaging and backreaction and on the effect of inhomogeneities on cosmic evolution and observations.
\end{abstract}
\submitto{\CQG}
\maketitle

\section{Introduction}\label{intro}

The well known Szekeres class of exact solutions do not admit (in general) isometry groups. They are subdivided \cite{Plebanski,Krasinski} in two classes: class I and II, with each class itself subdivided into three subclasses: quasi-spherical, quasi-flat and quasi-hyperbolic, depending on their limiting symmetric solution admitting a group of three Killing vectors acting on 2-dimensional orbits (spherical, plane and pseudo--spherical symmetry). 

The quasi--spherical Szekeres models of class I (Szekeres-I) with a dust source (with zero and nonzero $\Lambda$) have been regarded as the most suitable for cosmological applications and thus have been widely used as models of cosmic structures generalizing the popular Lema\^\i tre--Tolman--Bondi (LTB) models (their particular spherically symmetric sub--case). Szekeres models with a dust source (class I and II) introduce in all covariant scalars an extra degree of freedom in the form of a dipole (see detailed discussion in \cite{Plebanski,Krasinski}). In the case of quasi-spherical models of class I this dipole is superposed to the monopole of spherical symmetry \cite{Plebanski}, thus allowing for the construction of models of more than one structure in an FLRW background, typically a central over--density or void, surrounded by elongated  wall--like structures or spheroidal ``pancakes'' (either over--densities or voids) marked by the dipole orientation, all of which provides a much better approach to cosmic structures \cite{Sussbol}. In particular, it is possible to device elaborated networks of structures placed at chosen  locations as part of the setting up of initial conditions \cite{SussDel} that can provide a good coarse grained description of our cosmography at scales of 100 Mpc. 

While an FLRW background emerges naturally in Szekeres-I, the FLRW limit of Szekeres-II models is much more contrived with a more natural homogeneous limit being the Kantowski--Sachs models. Thus, in contrast with the widespread usage of Szekeres--I models,  Szekeres--II models have not received much attention in theoretical studies of the effects of inhomogeneity or in cosmological applications. Known articles involve their treatment as exact dust perturbations to compute the growth factor  \cite{Kasai1, Ishak1,Ishak2}, self consistency of perfect fluid thermodynamics \cite{SusQue,Coll} and more recently Delgado and Buchert have used Szekeres--II dust models as test case spacetimes to probe a formalism to obtain a relativistic generalization of the Zeldovich approximation in terms of spacetime averaging \cite{DelBuch}. In fact, these authors present a periodic ``lattice model'' that is a simplified particular case of the pancake models we are discussing in the present paper.  

However, while in a comoving frame class I models are Petrov type D with vanishing magnetic Weyl tensor, Szekeres--II models in full generality are Petrov type I and have nonzero magnetic Weyl tensor, and thus are compatible with a more general energy--momentum tensor including energy flux, all of which makes them good candidates to describe a wider variety of sources. We believe that these models have a good unexplored application potential in cosmology. 

In this article we present a novel class of new and interesting toy models based on Szekeres-II models but considering their fully general energy--momentum tensor, which admits non--trivial pressure gradients, anisotropic stresses and energy flux. Besides the appealing possibility of these extra degrees of freedom, we show that the free functions in their quasi--plane sub--class allows for a smooth matching with any spatially flat homogenous and isotropic model that can be described by the Robertson--Walker metric: FLRW models and also de Sitter, anti de Sitter and Minkowski spacetimes. 

The matchings of the type described above can be performed at an arbitrary number of hypersurfaces (diffeomorphic to time evolved 2--dimensional flat space), leading to compound configurations made of a series of (possibly different) Szekeres--II sections separated by regions of the same FLRW (or Minkowski or de Sitter or anti de Sitter) model, thus providing an isotropic and homogeneous background that was not thought possible for these models. These are a sort of  ``pancake'' configurations that can be regarded as quasi--flat (they are not spatially flat) analogues to the spherically symmetric ``Swiss-cheese models'' \cite{Ellis} or slab--like versions of lattice universes \cite{Larena1,Clifton,DelBuch}. 

The compatibility of Szekeres--II models with a more general general energy--momentum tensor opens the potential to describe a wider variety of matter--energy sources, such as dissipative fluids and fluid mixtures with non--comoving 4--velocities with non--trivial peculiar velocities, as well as fluid mixtures with scalar and magnetic fields or gravitational waves. Besides cosmological applications, these configurations can serve to probe and experiment with theoretical formalisms, such as averaging and the backreaction effect of local inhomogeneities in cosmic observations evolution \cite{Cliftonback}.

The section by section description of the paper is as follows. In section \ref{sec:SzekII} presents a brief introduction to Szekeres-II and FLRW spacetimes. In section \ref{sec:junc} we show that junction conditions hold for a smooth matching between the quasi--flat subclass along an arbitrary countable number of suitable hypersurfaces. In section \ref{sec:example} we present simple examples involving various examples of Szekeres-II pancake regions ``sandwiched'' between FLRW and de Sitter regions. In section \ref{perturbations} we show that the Szekeres-II pancake regions can be rigorously considered as exact covariant perturbations on an FLRW background defined by the smooth matching. Finally in section \ref{sec:disc} we discuss the obtained results and suggest future applications and extensions.

\section{General Szekeres-II models}\label{sec:SzekII}
The metric element characterizing Szekeres-II solution is
\begin{equation}
ds^2=-dt^2+S^2(t)\left[X^2 dw^2+\frac{dx^2+dy^2}{f^2}\right],\qquad f=1+\frac{k\,[x^2+y^2]}{4},\label{Szmetric}
\end{equation}
where $X=X(t,x^i)$ with $x^i=w,x,y$. This metric identifies a canonical orthonormal tetrad $e_{(\alpha)}^a$ such that $g_{ab}\,e_{(\alpha)}^a e_{(\beta)}^b =\eta_{(\alpha)(\beta)}$:
\begin{equation}
e^a_{(0)}=\delta^a_0, \quad e^a_{(w)}=\frac{1}{SX}\delta^a_w,\quad e^a_{(x)}=\frac{f}{S}\delta^a_x,\quad e^a_{(y)}=\frac{S}{f}\delta^a_y,\label{tetrad}
\end{equation}
and is compatible with a quite general energy-momentum tensor in a comoving frame ($u^a=e^a_{(0)}=\delta^a_0$):
\begin{equation}
T^{ab} = (\rho + \Lambda) u^a u^b +  (p-\Lambda) h^{ab}+ \pi^{ab}+ 2 q^{(a} u^{b)}.
\end{equation}
where energy density, isotropic and anisotropic pressures and energy flux, $\rho,\,p,\,\pi^{ab} =[h^{(a}_ch^{b)}_d-\frac13h^{ab}h_{cd}]T^{cd}$ and $q^a=u^bT^a_b$ (with $h^{ab}=u^au^b+g^{ab}$), depend in general on the four coordinates $t,x^i$. The corresponding nonzero field equations $G^{ab}=\kappa T^{ab}$ for $\kappa =8\pi G/c^4$ are
\begin{eqnarray}
\fl\kappa \rho =\kappa\bar\rho-\frac{f^2(X_{,yy}+X_{,xx})}{S^2X}+\frac{2\dot{S}\dot{X}}{SX},\qquad \kappa\bar\rho = \frac{3\dot{S}^2}{S^2}+\frac{3k}{S^2}-3\kappa\Lambda,\label{eq:eqrho}\\
\fl\kappa p =\kappa\bar p+\frac{1}{3}\frac{f^2(X_{,yy}+X_{,xx})}{S^2X}-\frac{2}{3}\frac{S\ddot{X}+3\dot{S}\dot{X}}{SX},\qquad\kappa\bar p = -\frac{2\ddot S}{S}-\frac{\dot S^2}{S^2}-\frac{k}{3S^2}+\Lambda,\label{eq:eqp}\\
\fl\kappa \pi^{xx}  =\frac{1}{3}\frac{f^4(2X_{,yy}-X_{,xx})}{S^4X}-\frac{1}{2}\frac{f^3k(xX_{,x}-yX_{,y})}{S^4X}-\frac{1}{3}\frac{f^2(S^2\ddot{X}+3S\dot{S}\dot{X}-kX)}{S^4X},\label{eq:eqpixx}\\
\fl\kappa \pi^{yy} =-\frac{1}{3}\frac{f^4(X_{,yy}-2X_{,xx})}{S^4X}+\frac{1}{2}\frac{f^3k(xX_{,x}-yX_{,y})}{S^4X}-\frac{1}{3}\frac{f^2(S^2\ddot{X}+3S\dot{S}\dot{X}-kX)}{S^4X},\label{eq:eqpiyy}\\
\fl\kappa \pi^{ww} =-\frac{1}{3}\frac{f^2(X_{,yy}+X_{,xx})}{S^4X^3}-\frac{2}{3}\frac{k}{S^4X^2}+\frac{2}{3}\frac{S\ddot{X}+3\dot{S}\dot{X}}{S^3X^3},\label{eq:eqpizz}\\
\fl\kappa \pi^{xy}=-\frac{f^3\,\left(fX\right)_{,xy}}{S^4X},\label{eq:eqpixy}\\
\fl\kappa q^x=\frac{f^2\dot{X}_{,x}}{S^2X},\label{eq:eqqx}\\
\fl\kappa q^y=\frac{f^2\dot{X}_{,y}}{S^2X}.\label{eq:eqqy}
\end{eqnarray}
where $\dot A=u^a A_{,a}$ for every function $A$. In a comoving frame the 4--acceleration  and vorticity tensor vanish, the nonzero kinematic parameters are then the expansion scalar $\Theta=\bar\nabla_a u^a$ and shear tensor $\sigma_{ab}=\tilde\nabla_{(a}u_{;b)}-(\Theta/3)h_{ab}$ given by
\begin{eqnarray}
 \Theta=\frac{\dot X}{ X}+\frac{3\dot{S}}{S},\qquad
 \sigma^a_b = \sigma \,\xi^a_b,\qquad \sigma=-\frac{\dot{X}}{3X},\label{eq:eqThSig}
\end{eqnarray}
where $\xi^a_b = h^a_b-3\delta^a_w\delta^w_b = \textrm{diag}[0, -2, 1, 1]$ and the spatial derivative projections are $\tilde\nabla_au^a=h_a^b u^a\,_{;b}$ and $\tilde\nabla_{(a}u_{;b)}=h_{(a}^ch_{b)}^d u_{c;d}$. 

To get an idea of the anisotropic evolution of comoving observers we compute from the expression for the symmetric expansion tensor $\Theta_{ab}=\bar\nabla_{a}u_{b}$ and its three eigenvalues $\Theta^a_b = \lambda_{(i)}\delta^a_b$: 
\begin{eqnarray} 
\lambda_{(1)}=\Theta^z_z =\frac{\dot S}{S}+\frac{\dot X}{X},\quad \lambda_{(2)}=\lambda_{(3)}=\Theta^x_x=\Theta^y_y =\frac{\dot S}{S},
 \end{eqnarray}  
so that kinematic anisotropy is clearly identified by the fact that the local expansion of fundamental observers along the principal directions $\lambda_{(2)}=\lambda_{(3)}$ along $e^a_{(x)}$ and $e^a_{(y)}$ is distinct from that of $\lambda_{(1)}$ along $e^a_{(w)}$. Another indicator of anisotropy comes from the local rate of change of redshift $z$ along a null geodesic segment \cite{Ellis}
\begin{equation}\label{eq:eqz}
\frac{d z}{z}=\left[\frac{1}{3}\Theta+\sigma_{ab}k^ak^b\right]\,d\vartheta,
\end{equation}
where $k^a$ is a null vector parametrized by the affine parameter $\vartheta$. Redshift from local observations are isotropically distributed only if $\sigma_{ab}=0$ (or $\dot X=0$). However, given the availability of extra degrees of freedom, the challenge is to constraint the inherent anisotropy of the models to limits set by observations.        

Szekeres--II models do not admit isometries (in general) but reduce to axial, spherical, flat and pseudo-spherical symmetry in suitable limits. The 2--surfaces marked by constant $t$ and $w$ have constant curvature that can be zero ($k=0,\,\,f=1$), positive ($k=1$) or negative ($k=-1$) respectively. Szekeres type-II solutions with $q_a\neq 0$ are Petrov type I, while solutions with $q_a=0$ (whether class I or II) are Petrov type D. The hypersurfaces of constant $t$ are conformally flat and the curve $\mathcal{C}(z)=[t_0,z,0,0]$ (with $x=0,y=0$) for an arbitrary fixed $t=t_0$ is a spacelike geodesic, whose tangent vector is a Killing vector of the 3--metric $h_{ab}$.

The momentum balance equations $\nabla_b T^{ab}=0$ are given by
\begin{eqnarray}
\dot\rho+(\rho+p)\Theta +\sigma_{ab}\pi^{ab}+\tilde\nabla_a q^a=0,\label{mombal1}\\
h_a^b\dot q_b+\frac43\Theta q_a+\tilde\nabla_b\pi^{ab}+\sigma_{ab}q^b+\tilde\nabla_a p=0,\label{mombal2}
\end{eqnarray}
while the Raychaudhuri and the constraints that define the energy flux $q_a$ the electric, magnetic Weyl tensor $H_{ab}$ are
\begin{eqnarray}
\dot\Theta=-\frac{\Theta^2}{3}-\frac{\kappa}{2}[\rho+3(p-\Lambda)]-\sigma_{ab}\sigma^{ab},\label{Raych}\\
\kappa q_a = \frac23\tilde\nabla_a\Theta+\tilde\nabla_b\sigma^{ab},\qquad H_{ab}=\hbox{curl}\,\sigma_{ab},\label{qH}\\
E_{ab}=\frac{\Theta}{3}\sigma_{ab}-\sigma_{c\langle a}\sigma^c_{b\rangle}-\frac{\kappa}{2}\pi_{ab}-{}^{(3)}{\cal R}_{\langle ab\rangle}, \label{eW}
\end{eqnarray}
where ${}^{(3)}{\cal R}_{\langle ab\rangle}$ is the spatially symmetric trace free Ricci tensor of the hypersurfaces orthogonal to $u^a$ (constant $t$), $\dot q_b=u^c\nabla_c q_b$ and $\hbox{curl}\,\sigma=\eta_{cd(a}\tilde\nabla^c\sigma_{b)}^d$ with $\eta_{abc}=\eta_{abcd}u^d$ for the Levi--Civita volume form $\eta_{abcd}=-\sqrt{-g}\,\epsilon_{abcd}$ with $\epsilon_{abcd}$ the totally antisymmetric unit tensor.   

We will use FLRW models described by the Robertson--Walker metric in rectangular coordinates
\begin{equation}
ds^2=-dt^2+\frac{a^2(t)\,(dx^2+dy^2+dz^2)}{\tilde{f}^2},\qquad \tilde{f}=1+\frac{\tilde{k}[x^2+y^2+z^2]}{4},\label{RWmetric}
\end{equation}
It only admits a perfect fluid energy-momentum tensor $T^{ab}=(\tilde{\rho}+\Lambda) \tilde u^a\tilde u^b +(\tilde p-\Lambda)\tilde h^{ab}$ with $\tilde u^a=u^a$ and $\tilde h^{ab}=\tilde g^{ab}+u^au^b$,  leading to the field equations and expansion scalar
\begin{eqnarray}
\fl\frac{\kappa}{3} (\tilde{\rho}+\Lambda) =\frac{\dot{a}^2}{a^2}+\frac{\tilde{k}}{a^2},\qquad 
\kappa (\tilde{p}-\Lambda) =-\frac{2\ddot a}{a}-\frac{\dot{a}^2}{a^2}-\frac{\tilde k}{3a^2},\qquad \tilde{\Theta}=\frac{3\dot{a}}{a}\label{eqsFLRW}
\end{eqnarray}
with the shear tensor vanishing everywhere. Considering that $t$ is the common proper time of fundamental observers in FLRW and Szekeres-II models, it is interesting to consider, from (\ref{RWmetric}) and (\ref{Szmetric}), an identification between the FLRW scale factor, density and pressure $a(t),\,\tilde\rho,\,\tilde p$ in (\ref{eqsFLRW}) vs. the  Szekeres-II metric function $S(t)$ and the purely time dependent part of the density and pressure $\bar\rho,\,\bar p$ in the field equations (\ref{eq:eqrho})--(\ref{eq:eqp}).   

\section{Matching between Szekeres-II models and FLRW}\label{sec:junc}

The Darmois conditions \cite{Israel, Mars} for a smooth matching between the two spacetimes $({\cal M}_{(+)},g_{(+)})$ and $({\cal M}_{(-)},g_{(-)})$, such as Szekeres-II and FLRW described by (\ref{Szmetric}) and (\ref{RWmetric}), along a matching hypersurface $\Sigma(x^\alpha)=0$, are the continuity of the first and second fundamental forms at $\Sigma$
\begin{eqnarray}
[\gamma_{ab}]= \gamma_{ab}^{(+)}-\gamma_{ab}^{(-)}=0, \qquad \gamma_{ab}=g_{ab} +\epsilon\,n_a n_b,\\
\left[K_{ab}\right]= K_{ab}|_{\Sigma_{+}}-K_{ab}|_{\Sigma_{-}}=0,\qquad K_{ab}=-n_{a;b}, \end{eqnarray}
where $\gamma_{ab}^{(\pm)}=\gamma_{ab}|_{\Sigma_{(\pm)}}=\lim_{z\to z_0^\pm}\gamma_{ab}$ (same for $K_{ab}$) and $n_a$ is the unit normal to $\Sigma$, so that $\epsilon =1,-1$ if the vectors tangent to $\Sigma$ are (respectively) timelike or specelike and ${}|_{\Sigma_{(\pm)}}$ denotes evaluation at $\Sigma$. Given the identification of coordinates $(t,x^i)$ and orthonormal tetrads in (\ref{Szmetric}) and (\ref{RWmetric}), we choose as $\Sigma=0$ the equation $w-w_0=0$ where $w_0$ is an arbitrary constant. We have then $n_a = e_a^{(w)}=S\,X\,\delta_a^w$ and the induced metric $\gamma_{ab}$ is parametrized by the coordinates $x^\alpha=[t,w_0,x,y]$ with $w_0$ an arbitrary fixed value of $w$, choosing $(+)$ and $(-)$ as Szekeres-II and FLRW we have
\begin{eqnarray}\label{1junc} \fl \gamma_{\alpha\beta}=e_{(\alpha)}^{a}e_{(\beta)}^{b}g_{ab}^{(\pm)},\quad \gamma_{tt}^{(+)}=\gamma_{tt}^{(-)}=-1, \quad \gamma_{xx}^{(+)}=\gamma_{yy}^{(+)}=\frac{S^2}{f^2}, \quad \gamma_{xx}^{(-)}=\gamma_{yy}^{(-)}=\frac{a^2}{\tilde f^2(w_0)}.\nonumber\\\end{eqnarray}
The components of the second fundamental form (extrinsic curvature of $\Sigma$) are
\begin{eqnarray}\label{21junc}
\fl K_{tw}^{(+)}= \left(X\dot{S} + \dot{X} S\right)^{(+)}, \;\;\; K^{(+)}_{tx} = \left(SX_{,x}\right)^{(+)}, \;\;\; K^{(+)}_{ty}= \left(SX_{,y}\right)^{(+)},\\
\fl K^{(-)}_{tw}=\dot{a}.\label{22junc}
\end{eqnarray}
Combining (\ref{1junc}) and (\ref{21junc})--(\ref{22junc}) implies that all derivatives of $X$ along the direction of tangent vectors to $\Sigma$ must vanish at $w=w_0$:
\begin{eqnarray} X^{(+)}=1,\quad (X_{,x})^{(+)}=(X_{,y})^{(+)}=(\dot X)^{(+)}=(\ddot X)^{(+)}=0,\label{31junc}\\
 S(t)=a(t),\qquad k=\tilde k=0 \quad\Rightarrow\quad f=\tilde f=1,\label{3junc}\end{eqnarray} 
so that the matching at $\Sigma$ given by $w=w_0$ is only possible between a quasi--plane Szekeres-II model and a spatially flat FLRW model. These matching conditions also require continuity of $T^a_bn^b=T^z_z/(SX)$: 
\begin{equation} \kappa(\bar p-\Lambda)= -\frac{2\ddot S}{S}-\frac{\dot S^2}{S^2} =  -\frac{2\ddot a}{a}-\frac{\dot a^2}{a^2} = \kappa (\tilde p-\Lambda)\end{equation} 
so that $S(t)$ must satisfy the same spatially flat ($\tilde k=0$) evolution equations as the FLRW scale factor $a(t)$ (\ref{eqsFLRW}), which fully identifies the FLRW density and pressure $\tilde\rho,\,\tilde p$ with the purely time dependent parts, $\bar\rho,\,\bar p$, of the quasi--plane ($k=0$) Szekeres-II density and pressure in (\ref{eq:eqrho})--(\ref{eq:eqp})). Notice that (\ref{1junc})--(\ref{3junc}) hold at $w=w_0$. Conditions  (\ref{1junc})--(\ref{3junc}) and also imply continuity of $\rho, p,\pi^{ab}, q_a$ at $\Sigma$ (the right hand side of (\ref{eq:eqpixx})-(\ref{eq:eqqy}) vanish at $\Sigma$).

Pending on the free functions that depend on $w$, the matching between quasi--flat Szekeres-II models and spatially flat FLRW spacetimes that we have described can be performed along an arbitrary number of hypersurfaces marked by constant $w$. The free parameters of $X$ in each Szekeres--II region would have to fulfill (\ref{31junc}) at two (or pairs of) fixed values of $w$. The resulting ``pancake'' configuration is a collection of Szekeres--II regions smoothly matched to a given FLRW spacetime. Notice that (\ref{22junc})--(\ref{3junc}) imply that all the Szekeres--II regions must be matched to the same FLRW model, which acts then as a background, while the Szekeres--II regions can be characterized by any source in which the free parameters can be set up to fulfill the matching conditions (\ref{31junc})--(\ref{3junc}). These type of configurations provide nice toy models to probe the effects of cosmological inhomogeneities.

\section{Particular solutions}

Given an assumption on the sources, it is not possible to know if the function $X$ can satisfy (\ref{1junc})--(\ref{3junc}) at an arbitrary $w=w_0$ without having a solution, analytic or numerical, of the field equations. However, more information can be obtained for an important particular case: if $f=1$ and the components $\pi^{xy}$ of the anisotropic pressure in equation (\ref{eq:eqpixy}) vanishes everywhere, we have $X_{,xy}=0$ whose general solution
is
\begin{equation} X= F(t,w)+\Phi(t,w,x)+\Psi(t,w,y), \label{solX1},\end{equation}
where $F,\,\Phi$ and $\Psi$ are entirely arbitrary. A simple way to make (\ref{solX1}) compatible with a matching with FLRW at $w=w_0$ is to assume that $\Phi,\,\Psi$ are separable in various forms, for example:
\begin{equation} X= F(t,w)+\sum_{j=1}^n N_j(t,w)\phi_A(w,x)+Q_j(t,w) \psi_j(z,y), \label{solX2},\end{equation}
so that (given a solution for a specific source) fulfillment of (\ref{1junc})--(\ref{22junc}) can be achieved (for example) by demanding the boundary conditions $F(t,w_0)=1$ and $\phi_j(x,w_0)=\psi_j(y,w_0)=0$. By inserting (\ref{solX2}) in (\ref{eq:eqrho})--(\ref{eq:eqqy}) we can identify various particular cases:
\begin{itemize} 
\item Zero energy flux ($q_a=0$) with nonzero anisotropic pressure ($\pi_{ab}\ne 0$): $\dot N_j(t,w)=\dot Q_j(t,w)=0$ for all $j$, leading to
\begin{equation} X= F(t,w)+\sum_{j=1}^n \bar\phi_j(w,x)+\bar\psi_j(w,y), \end{equation}
where $\bar\phi_j(w,x) =\phi_j(w,x) N_j(w)$ and $\bar\psi_j(w,y)=\psi_j(w,y) Q_j(w)$.
\item Zero anisotropic pressure ($\pi_{ab}=0$) nonzero energy flux ($q_a\ne 0$). These solutions follow from (\ref{solX2}) with: 
\begin{eqnarray}\dot N_j=0, \quad Q_j=Q_1=Q(t,w),\quad X=F+A+Q\,B,\label{noPIyesQ1}\\
 \bar\phi_j=A(w,x,y)=\alpha_0(w)+\alpha_1(w)x+\alpha_2(w)y+\alpha_3(w)(x^2+y^2), \label{noPIyesQ2}\\
 \bar\psi_j=B(w,x,y)=\beta_0(w)+\beta_1(w)x+\beta_2(w)y+\beta_3(w)(x^2+y^2), \label{noPIyesQ3}\end{eqnarray}
while $F$ and $Q$ must satisfy the coupled linear differential equations
\begin{eqnarray} \ddot F+\frac{3\dot S}{S}\dot F-\frac{2(\beta_3\,Q+\alpha_3)}{S^2}=0,\label{ecdF}\\ 
\ddot Q +\frac{3\dot S}{S}\dot Q=0\,\,\Rightarrow\,\, Q=c_1(w)+c_0(w)\int{\frac{dt}{S^3}}\label{ecdQ}\end{eqnarray}
It is straightforward to verify that (\ref{noPIyesQ1})--(\ref{ecdQ}) lead to $p=\bar p(t)$ with $k=0$ in (\ref{eq:eqp}) but $\rho=\rho(t,x^i)$. The particular case of dust ($p=\bar p(t)=0$) is the solution found by Goode \cite{Goode} (though Goode assumed that $c_1=0,\,c_0 = \hbox{const.}$ and $Q=Q(t)$).
\item Perfect fluid: as in (\ref{noPIyesQ1})--(\ref{ecdF}) with $\dot Q = 0$, hence (\ref{ecdQ}) is redundant and $X=F+A$. In general, perfect fluid solutions are characterized by $\bar p=\bar p(t)\ne 0$ with $\rho=\rho(t,x^i)$, some of these solutions have a mathematically consistent thermodynamical interpretation \cite{SusQue,Coll}. Dust solutions examined in \cite{Kasai1,Bruni,Ishak1,Ishak2} follow by setting $p=\bar p(t)=0$ in (\ref{eq:eqp}).   
\end{itemize}
Being a second order linear PDE, it is evident that a solution of (\ref{ecdF}) will have the form $F=\varphi_{+}(w)F_{+}(t,w)+\varphi_{-}(w)F_{-}(t,w)$ with $\varphi_{\pm}(w)$ integration constants. Therefore, in all the cases summarized above the free parameters depending on $w$ can be fixed so that (\ref{1junc})--(\ref{3junc}) hold at an arbitrary $w=w_0$ and $S(t)$ can be identified with the scale factor $a(t)$ of a given FLRW spacetime. 

\begin{figure}
\centering
%\begin{minipage}{.45\textwidth}
  \centering
  \includegraphics[width=1\linewidth]{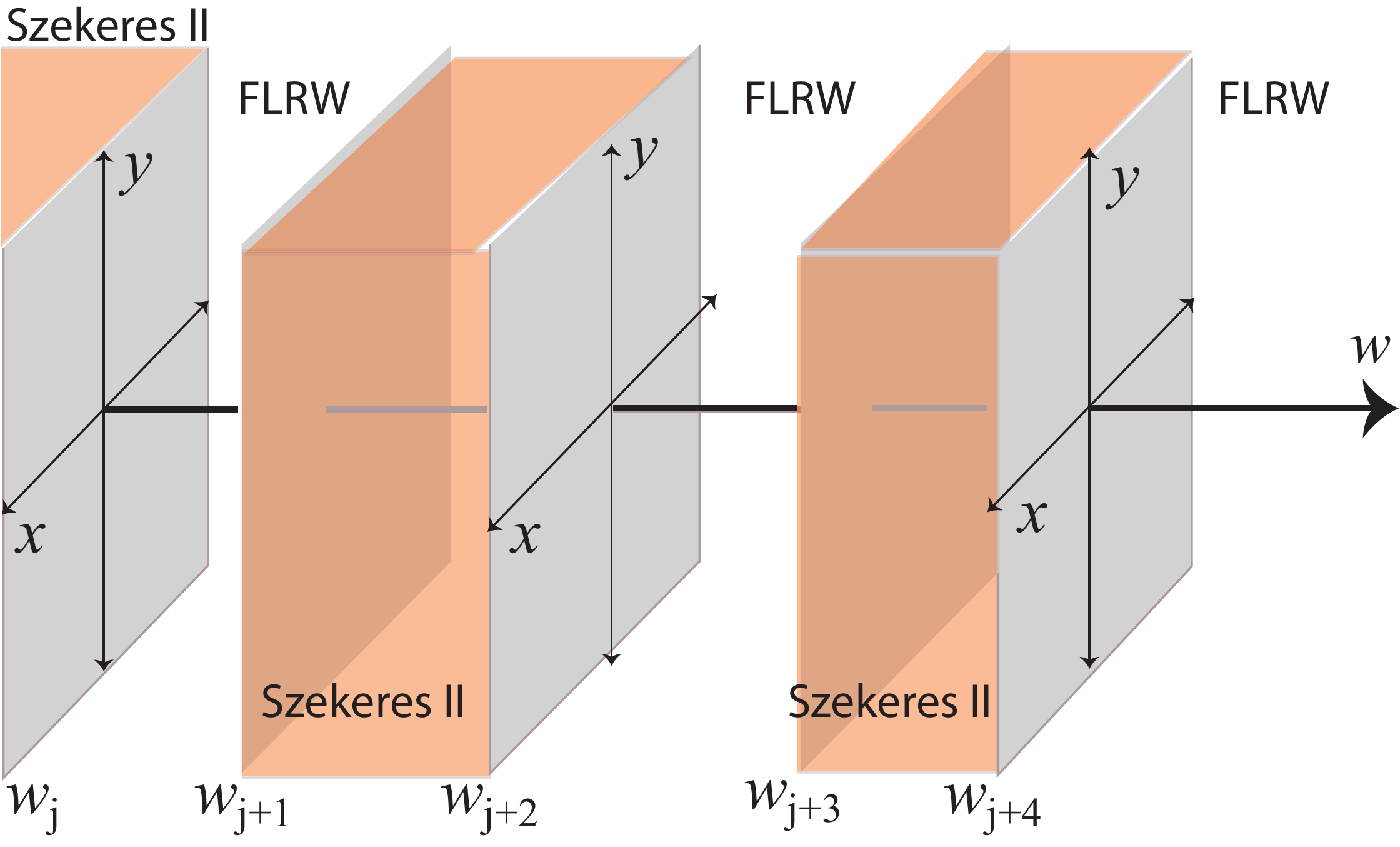}
  \caption{Matchings of different Szekeres--II regions with an FLRW spacetime. The figure shows a $t$ constant hypersurface, with two Szekeres II regions located between fixed values of the coordinate $w$ in the ranges  $w_{j+1}<w<w_{j+2}$,\,\,$w_{j+3}<w<w_{j+4}$ and FLRW regions occupying the remaining range of $w$. The curve $x=y=0$ for arbitrary $w$ (the $w$ axis) is a geodesic and a Killing vector of the 3--metric. }\label{fig1}
%\end{minipage}%
 %\hspace{0.05\linewidth}
%\begin{minipage}{.45\textwidth}
 % \centering
%  \includegraphics[width=1\linewidth]{Images/Omrho0z}
%  \caption{Average $\Omega_\rho$ at constant $t,r$ for different values of $r$.}\label{fig:Omz0}
%\end{minipage}
\end{figure} 
            
\section{Examples of Szekeres-II sections matched to FLRW sections}\label{sec:example}

We examine in this section the multiple matching configurations that can be constructed for various Szekeres--II and FLRW regions.  

\subsection{Szekeres--II dust and $\Lambda$CDM}

We consider the case $p=Q=0$ in (\ref{noPIyesQ1})--(\ref{ecdF}) with the corresponding FLRW spacetimes being sections of a $\Lambda$CDM model whose Friedman equation (\ref{eqsFLRW}) (with $\tilde k=0$) and associated equation (\ref{ecdF}) have the following solutions 
\begin{eqnarray}
\fl S(t) = \lambda_1\,\sinh^{2/3}\left(\lambda_2\,\tau\right),\qquad \bar{\cal H}\equiv\frac{\bar H}{\bar H_0}=\frac{\dot S/S}{\bar H_0}=\frac{2\lambda_2 \bar H_0}{3}\coth\left(\lambda_2\,\tau\right),\label{ecS1}\\
\fl F(t,w)=\varphi_1(w)+\varphi_2(w)\,F_1(\tau)+\frac{2\alpha_3(w)}{\lambda_1^2}\,F_2(\tau),\label{ecF1},\\
\fl F_1(\tau) = -\tau -\frac{1}{3\lambda_2}\left[2\hbox{coth}\left(\frac32\lambda_2\tau\right)+\ln\left(
\frac{\hbox{coth}(\frac32\lambda_2\tau)-1}{\hbox{coth}(\frac32\lambda_2\tau)+1}\right)\right],\\
\fl F_2(\tau) =,\int{\frac{ \int{\sinh^{2/3}(\lambda_2\,\tau)d\tau}}{\sinh^2(\lambda_2\,\tau)}\,d\tau},
\end{eqnarray}
where $\tau= \bar H_0t,\,\,\lambda_1=(\Omega_0^m/\Omega_0^\Lambda)^{1/3}$ and $\lambda_2=\sqrt{\Omega_0^\Lambda}$, with $\bar H_0$ the present day Hubble factor and $\Omega_0^m=\kappa\rho_0/(3\bar H_0^2),\,\,\Omega_0^\Lambda=\kappa\Lambda/(3\bar H_0^2)$.

A multiple ``pancake'' configuration made of Szekeres-II dust and $\Lambda$CDM model can be constructed by setting up smooth matchings along multiple hypersurfaces marked by $n$ arbitrary distinct fixed values $w=w_0^\gamma$ with $\gamma=1\dots n$. The simplest way to fulfill the matching conditions (\ref{1junc})--(\ref{3junc}) is to assume for both cases above that 
\begin{equation}S(t)=a(t),\qquad \varphi_1(w_0^\gamma)=1,\quad \varphi_2(w_0^\gamma)=\alpha_3(w_0^\gamma)=0,\label{match1}\end{equation}
holds for all $w_0^\gamma$. From (\ref{eq:eqrho}), (\ref{eq:eqThSig}) and (\ref{noPIyesQ2}) the normalized dimensionless density and Hubble scalar of the Szekeres--II regions are 
\begin{eqnarray}\hat\Omega\equiv \frac{\kappa\rho}{3 \bar H_0^2}=\bar\Omega+\frac{1}{F+A}\left[\frac{2\bar H}{3\bar H_0}\frac{\partial F}{\partial \tau}-\frac{4\alpha_3}{S^2}\right],\label{rhoLCDM}\\
\hat {\cal H}\equiv \frac{\Theta}{3\bar H_0} = \bar{\cal H}+\frac{1}{3(F+A)}\frac{\partial F}{\partial \tau},\qquad \frac{\sigma}{\bar H_0} = -\frac{1}{3(F+A)}\frac{\partial F}{\partial \tau},\label{HLCDM}\end{eqnarray}
where $A,\,S$ and $F$ are given by (\ref{noPIyesQ2}) and (\ref{ecS1})--(\ref{ecF1dS}) and 
\begin{equation}\bar\Omega \equiv \frac{\kappa\bar\rho}{3\bar H_0^2}= \frac{\Omega_0^m}{S^3}+\Omega_0^\Lambda,\qquad \bar{\cal H} \equiv \frac{\dot a/a}{\bar H_0^2}=\sqrt{\bar\Omega}, \end{equation}
are the dimensionless density and Hubble scalar of the $\Lambda$CDM regions.

\subsection{Dust in a de Sitter background}

Proceeding as in the case before, we now have $\tilde \rho=k=0$ in (\ref{eq:eqrho}) and since the matching requires $S(t)=a(t)$ we have $\bar\rho=\bar k =0$ in (\ref{eqsFLRW}). The corresponding functions $S$ and $F$ and parameters of the de Sitter regions are
\begin{eqnarray}
S(\tau) = S_0\exp \left(\lambda_2\,\tau\right),\qquad \bar{\cal H}= \lambda_2,\qquad \bar\Omega=\lambda_2^2=\Omega_0^\Lambda,\label{ecS1dS}\\
F = \varphi_1(w) -\frac{\varphi_2(w)}{3\lambda_2}\exp\left(-3\lambda_2\,\tau\right)-\frac{2\alpha_3(w)}{\lambda_2^2}\exp\left(-2\lambda_2\,\tau\right),\label{ecF1dS}
\end{eqnarray}   
where $\bar H,\,\tau$ and $\lambda_2$ were defined above and $S_0=S(\tau_0)$. The dimensionless energy density is given by 
\begin{equation}
\fl \hat\Omega =  \Omega_0^\Lambda\left[1 + \frac{2}{3}\frac{(4S_0^2 - 6)\alpha_3\exp(-2\sqrt{\Omega_0^\Lambda}\tau) + S_0^2\Omega_0^\Lambda\varphi_2\exp(-3\sqrt{\Omega_0^\Lambda}\tau)}{S_0^2\,[F(\tau, w) + A(w, x, y)]}\right],\label{ecrhodS}
\end{equation}
where $F$ and $A$ are given by (\ref{ecF1dS}) and (\ref{noPIyesQ2}) and we have substituted $\lambda_2^2=\Omega_0^\Lambda$. To explore the properties of these solutions, we choose the simple particular case  $S_0=\varphi_1=1$,\, $\alpha_1=\alpha_2=\varphi_2=0$, leading to  
\begin{equation}\hat\Omega = \Omega_0^\Lambda\,\frac{(1+\alpha_0+\alpha_3 r^2)\Omega_0^\Lambda\,e^{2\sqrt{\Omega_0^\Lambda}\tau}-\frac{10}{3}\alpha_3}{(1+\alpha_0+\alpha_3 r^2)\Omega_0^\Lambda\,e^{2\sqrt{\Omega_0^\Lambda}\tau}-2\alpha_3},\label{ecrhodS}   
\end{equation}
where $\Omega_0^\Lambda$ can be selected to match the value of this parameter in a $\Lambda$CDM model, $r^2=x^2+y^2$ (we use polar coordinates since this case is axially symmetric with $r=0$ the axis of symmetry) and the functions $\alpha_0(w),\,\alpha_3(w)$ selected to comply with the matching with de Sitter spacetime (see further ahead).
The asymptotic limits of (\ref{ecrhodS}) are
\begin{eqnarray}\hat\Omega\to\Omega_0^\Lambda\quad \tau\to\infty\,\,\,\hbox{and}\,\,\,r\to\infty,\qquad 
\hat\Omega\to\frac53\Omega_0^\Lambda\quad \tau\to-\infty,\label{asympt}
 \end{eqnarray}
so that the Szekeres--II region has two branches with positive dust density separated by a singularity and a region with negative density. The curvature singularity is marked by a zero of the denominator in (\ref{ecrhodS}), but the zero of the numerator is located to the future of the singularity (see figure 2). Hence, for $\tau$ immediately in the future of the singularity the numerator is negative marking a region close to the singularity with negative density so that $\hat\Omega\to -\infty$ at the singularity. As $\tau$ increases $\hat\Omega$ becomes positive reaching in the infinite future the asymptotic limit shown in (\ref{asympt}) (see figure 2b). In the past of the singularity $\hat\Omega$ is positive with $\hat\Omega\to\infty$ as the singularity is approached and the asymptotic limit in the infinite past shown in (\ref{asympt}). Notice from figure 2a how the locus of the singularity bends towards the infinite past as $w$ approaches the matching interface with de Sitter (which has no singularity).          
\begin{figure}
\centering
  \includegraphics[width=1\linewidth]{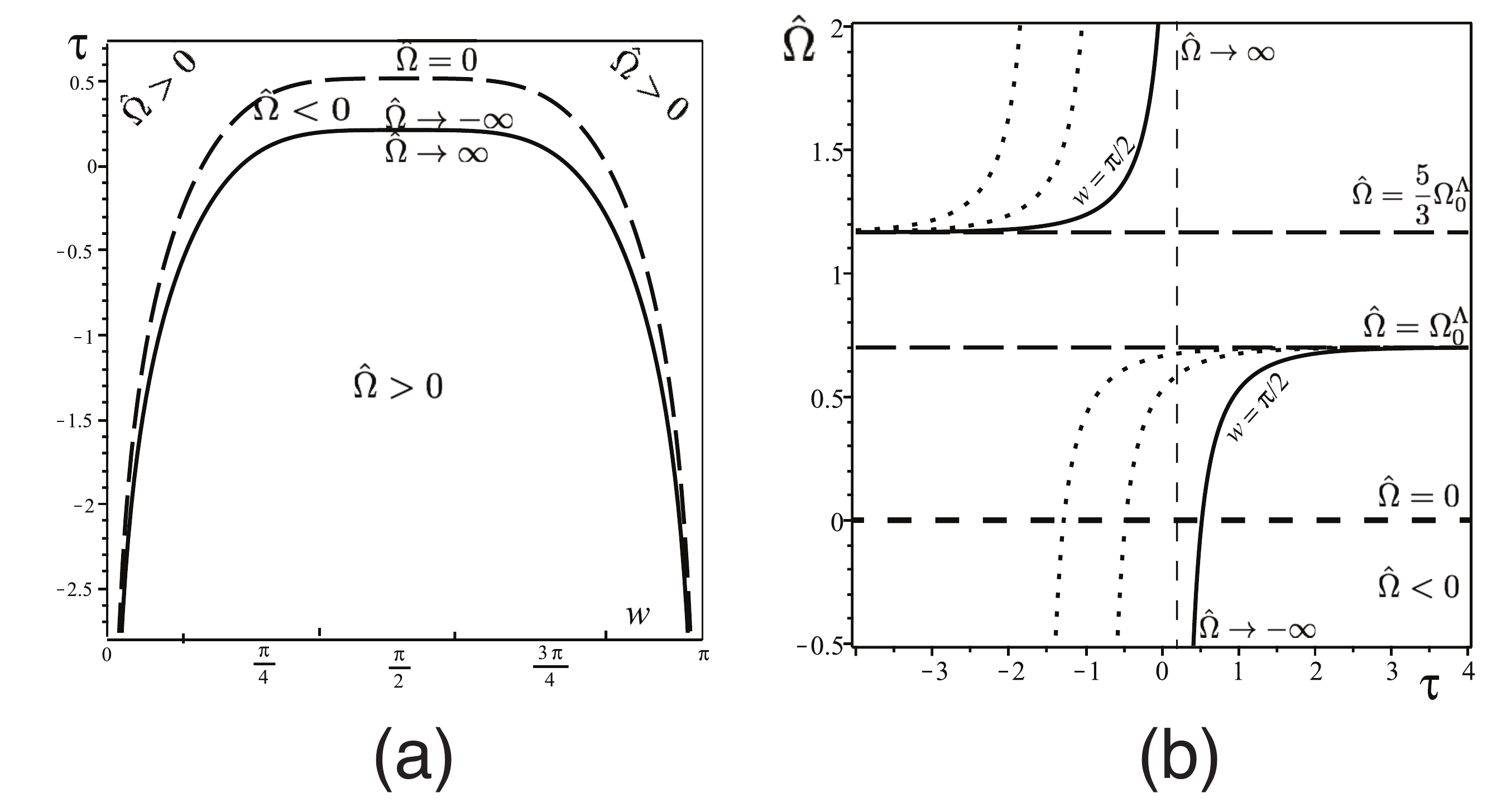}
  \caption{Normalized density $\hat\Omega$ for dust matched with de Sitter. Panel (a) displays $\hat\Omega$ at the axis $r=0$ for in the plane $[w,\tau]$. The two regions where $\hat\Omega>0$ are separated by a curvature singularity and a region with $\hat\Omega<0$. Panel (b) displays the time evolution of $\hat\Omega$ for $r=0$ and varios fixed values of $w$ (the thick curve is $w=\pi/2$. Notice how comoving observers reach $\hat\Omega\pm\to\infty$. The free functions used for the graphs are $\alpha_3=\sin^2 w$ and $\alpha_0=\sin^4 w$.}\label{fig:locusBB}
\end{figure} 

A multiple ``pancake'' configuration of $n$ smooth matchings between Szekeres--II dust and de Sitter along arbitrary $w=w_0^\gamma$ with $\gamma=1..n$ can be constructed as in the case of Szekeres--II and $\Lambda$CDM. The dust density is (\ref{ecrhodS}), the expansion scalar follows from (\ref{HLCDM}) with $S,\,F$ given by (\ref{ecS1dS})--(\ref{ecF1dS}), $A$ by (\ref{noPIyesQ2}) and setting $\Omega_0^m=0$ and $\Omega_0^\Lambda=\lambda_2^2$. Since the functions $\alpha_3(w),\,\varphi_1(w),\,\varphi_2(w)$ are not affected by the derivative $\dot F$, it is evident that (\ref{match1}) are sufficient to fulfill (\ref{1junc})--(\ref{3junc}) at all $w_0^\gamma$.

\subsection{Szekeres--II sections with energy flux as a non--comoving peculiar velocity field}

The ``pancake'' configurations constructed with smoothly matched Szekeres--II and FLRW sections can also accommodate the case when the source of the Szekeres--II sections is not a perfect fluid. As an example we consider here the case with nonzero energy flux and zero anisotropic pressure described by (\ref{noPIyesQ2})--(\ref{ecdQ}). In particular, as mentioned before, the dust subcase with zero isotropic pressure $\bar p(t)=0$ (and complying with (\ref{ecdQ})) is a generalization of the ``heat conducting dust'' solution found by Goode \cite{Goode} characterized by
\begin{equation} T^{ab}=\rho u^a u^b + 2q^{(a}u^{b)}.\label{GoodeTab}\end{equation}
However, instead of Goode's interpretation of $q^a$ as a heat conducting vector, we provide a wholly different and much more physically plausible interpretation in terms of non--relativistic peculiar velocities \cite{Ellisgen} of a non--comoving dust source that can be an adequate model for CDM. Non--comoving and comoving  4--velocities $\hat u^a$ and $u^a$ are related through a peculiar velocity field $v^a$ by a boost factor 
\begin{equation} u^a =\gamma\,\left[\hat u^a+v^a\right],\qquad \gamma = \frac{1}{\sqrt{1-v_av^a}},\quad v_au^a=0.\label{boost}\end{equation}
Substitution of (\ref{boost}) into (\ref{GoodeTab}) and considering non--relativistic peculiar velocities ({\it i.e.} $v^a/c\ll 1$ so that $\gamma = 1+O(v^2/c^2)$) we obtain up to linear terms in $v/c$ the energy momentum tensor (\ref{GoodeTab}) with $q_a=[0,0,q_x,q_y]$ given by (\ref{eq:eqqx})--(\ref{eq:eqqy}). Transformation of $q_x,\,q_y$ into $q_r,\,q_\theta$ into polar coordinates $(r,\theta)$ defined by $x=r\,\cos\theta,\,\,y=r\,\sin\theta$ yields
\begin{equation} \fl q_a = \rho v_a,\qquad q_r = \frac{c_0\,\left[\beta_1\cos\theta+\beta_2\sin\theta+2\beta_3\,r\right]}{S^5\left[F+A+QB\right]},\quad 
q_\theta = \frac{c_0\,r\,\left[-\beta_1\sin\theta+\beta_2\cos\theta\right]}{S^5\left[F+A+QB\right]},\label{qfoms}\end{equation}          
where the arbitrary $w$-dependent functions $c_0$ and $\beta_1,\beta_2,\beta_3$ in (\ref{noPIyesQ2}), as well as the functions $\alpha_1,\alpha_2,\alpha_3$ in (\ref{noPIyesQ3}), are chosen to comply with matching conditions (\ref{1junc})--(\ref{3junc}). Notice that $q_r\to 0$ as $r\to 0$, since for every fixed $w$ the curve $\beta_1\cos\theta+\beta_2\sin\theta\to 0$ as $r\to 0$ in the plane $[r,\theta]$ in polar coordinates. The functions $F$ and $Q$ are determined by (\ref{ecdF})--(\ref{ecdQ}) for a given choice of $S(t)=a(t)$ of the compatible dust FLRW spacetime to be matched. Notice that the assumption $|v^a|/c\ll 1$ does not imply that the gradients of $v_a$ are also small. We discuss this issue in Appendix A. 

Since $\Lambda=0$ in the energy--momentum tensor (\ref{GoodeTab}) that we are considering for the Szekeres--II regions, the matched FLRW spacetime cannot be a $\Lambda$CDM model compatible with observations (in such case (\ref{ecdF}) would need to be solved numerically). Hence, the matched FLRW spacetime must be the Einstein de Sitter model (spatially flat FLRW dust with $\Lambda=0$) for which analytic solutions of (\ref{ecdF}) are readily available. While the Einstein de Sitter model is not realistic, it serves the purpose of illustrating the ``pancake'' configurations (we examine in a separate paper the case of (\ref{GoodeTab}) with $\Lambda>0$ that admits matching with $\Lambda$CDM model).  The functions $S,\,Q$ and $F$ for an Einstein de Sitter matching are
\begin{eqnarray}
\fl S(\tau)&=a(\tau)=\left(\frac{3\tau}{2}\right)^{\frac{2}{3}},\label{ecSQ}\qquad 
\bar\Omega = \frac{4}{9\tau^2},\qquad \bar{\cal H}=\frac{2}{3\tau},\\
\fl F(\tau,w)&=\varphi_0-\frac{\varphi_1}{\tau}+\frac{(c_1\beta_3+\alpha_3)\,(12\tau)^{2/3}}{5}-\frac{12^{2/3}c_0\beta_3}{\tau^{1/3}},\qquad Q=c_1(w)+\frac{c_0(w)}{\tau},\label{ecFF}		
\end{eqnarray}
where $\tau=\bar H_0\,t$ and $\varphi_0,\varphi_1$ are two extra are free functions of $w$. 
The dimensionless density and Hubble scalar can be computed directly by inserting (\ref{noPIyesQ1})--(\ref{noPIyesQ3}), (\ref{ecSQ})--(\ref{ecFF}) in (\ref{eq:eqrho}) and (\ref{eq:eqThSig}) (with $f=1$) 
\begin{eqnarray}
\hat\Omega=\frac{4}{9\tau^2}\left[1+\frac{\gamma_1\tau^{5/3}+\gamma_2}{\gamma_3\tau^{2/3}-\gamma_1\tau^{5/3}-\gamma_4\tau-\gamma_2}\right],\label{eqhatOM}\\ 
\hat {\cal H} = \frac{2}{3\tau}\left[1 - \frac{\gamma_1\tau^{5/3} + \frac12\gamma_3\tau^{2/3} +\frac32\gamma_2}{\gamma_3\tau^{2/3}-\gamma_1\tau^{5/3}-\gamma_4\tau-\gamma_2}\right],\label{eqhatH}
\end{eqnarray}
with
\begin{eqnarray}
\gamma_1(w)=\alpha_3+c_1\beta_3,\quad \gamma_2(w,r,\theta) = 5(\varphi_1-c_0B),\\
 \gamma_3(w)=5c_0\beta_3,\quad \gamma_4(w,r,\theta)=5(\varphi_0+A+c_1B), 
\end{eqnarray}
where $A,\,B$ are given by (\ref{noPIyesQ2})--(\ref{noPIyesQ3}). The peculiar velocities follow from (\ref{qfoms}) and (\ref{eqhatOM})
\begin{eqnarray}
\fl v^r =  \frac{\left(\frac32\right)^{2/3}\,c_0\left[\beta_1\cos\theta+\beta_2\sin\theta+2\beta_3\,r\right]}{\tau\left[(12)^{2/3}c_0\beta_3-\gamma_4(w,r,\theta)\,\tau^{1/3}\right]},\quad
v^\theta = \frac{\left(\frac32\right)^{2/3}\,c_0\left[-\beta_1\sin\theta+\beta_2\cos\theta\right]}{r\,\tau\left[(12)^{2/3}c_0\beta_3-\gamma_4(w,r,\theta)\,\tau^{1/3}\right]},
\end{eqnarray} 
There are many possibilities to choose the free functions of $w$ in (\ref{noPIyesQ2})--(\ref{noPIyesQ3}) plus $c_0,\,c_1,\,\varphi_0,\varphi_1$ to comply with the matching conditions (\ref{21junc})--(\ref{3junc}). A particularly simple choice is to choose at every matching interface $\varphi_0(w_0^\gamma)=1$, with the remaining free functions vanishing at the interfaces. A convenient choice is $\varphi_0=\cos^2(\nu_0 w)$ with the rest of the free functions taking the form $\sin^2(\nu w)$ where the values of $\nu_0,\,\nu$ chosen to comply with the matching conditions. Evidently, the selection of the free functions must assure a positive density $\hat\Omega>0$, though the conditions to avoid shell crossings associated with $X=0$ for $S>0$ need to be tested on a case by case basis.      

\begin{figure}
    \centering
    \begin{subfigure}[b]{0.485\textwidth}
        \includegraphics[width=\textwidth]{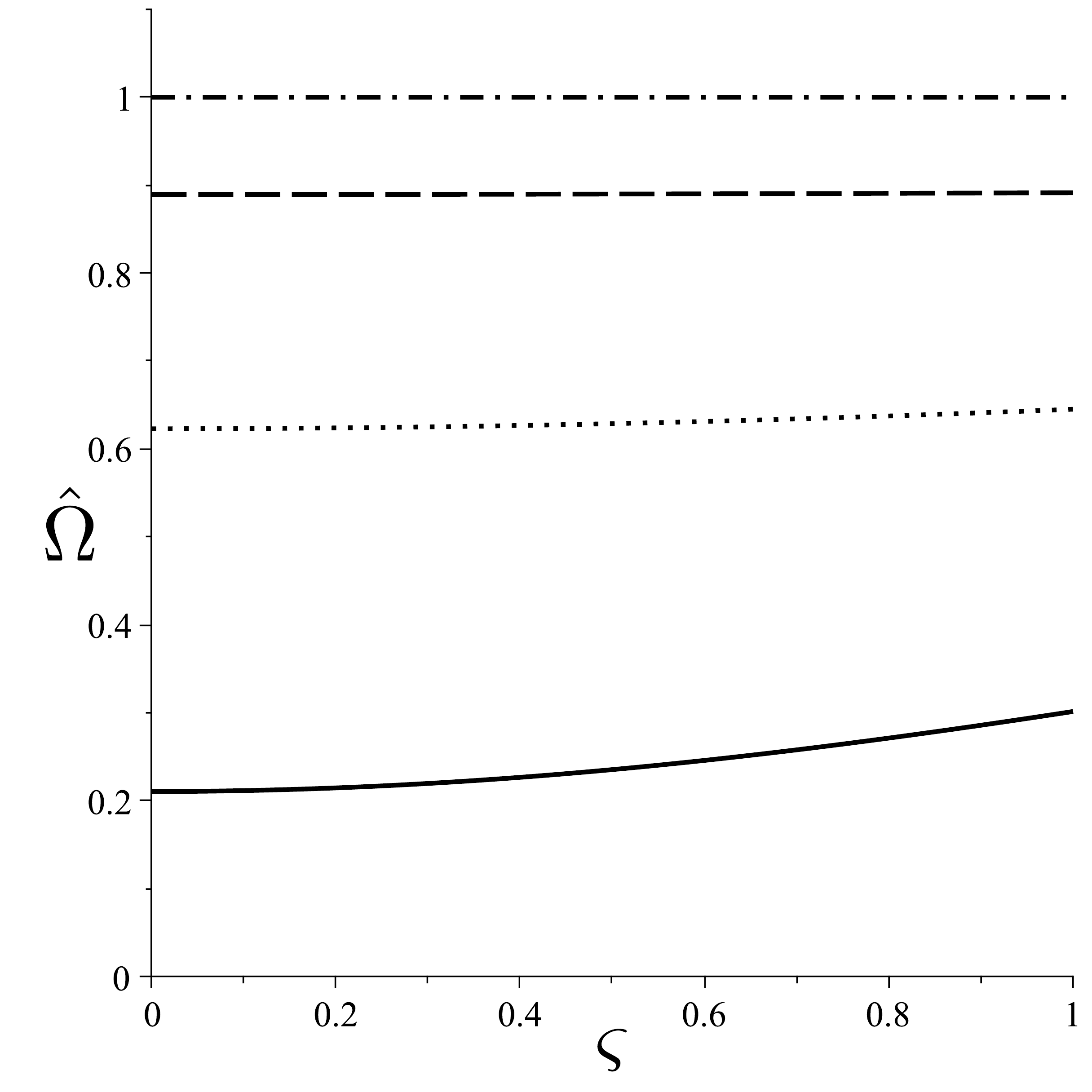}
        \centering
        \caption{}
        \label{fig:Omr0}
    \end{subfigure}
    ~ %add desired spacing between images, e. g. ~, \quad, \qquad, \hfill etc. 
    \begin{subfigure}[b]{0.485\textwidth}
        \includegraphics[width=\textwidth]{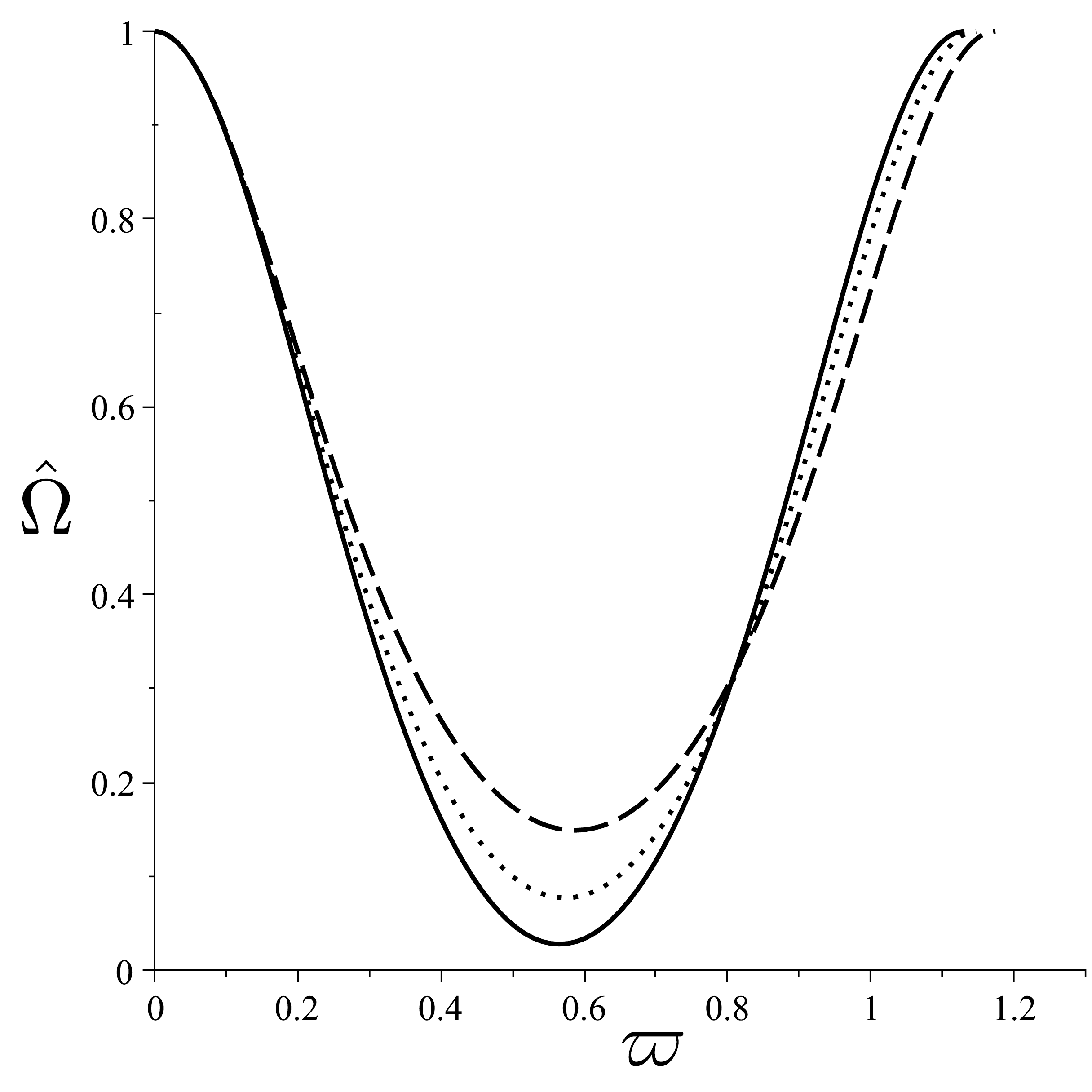}
        \centering
        \caption{}
        \label{fig:Omz0}
    \end{subfigure}
    \caption{Normalized density $\hat\Omega$ at present time, $\tau_0$, for dust matched with a Szekeres--II section with energy flux. Panel (a) displays $\hat\Omega$ vs. $\varsigma=cSr/\bar H_0$  for various values of $w$. Panel (b) displays $\hat\Omega$ vs. $\varpi=S\int X dw$ for various values of $r$.}\label{fig:OmSz}
\end{figure}

\begin{figure}
    \centering
    \begin{subfigure}[b]{0.485\textwidth}
        \includegraphics[width=\textwidth]{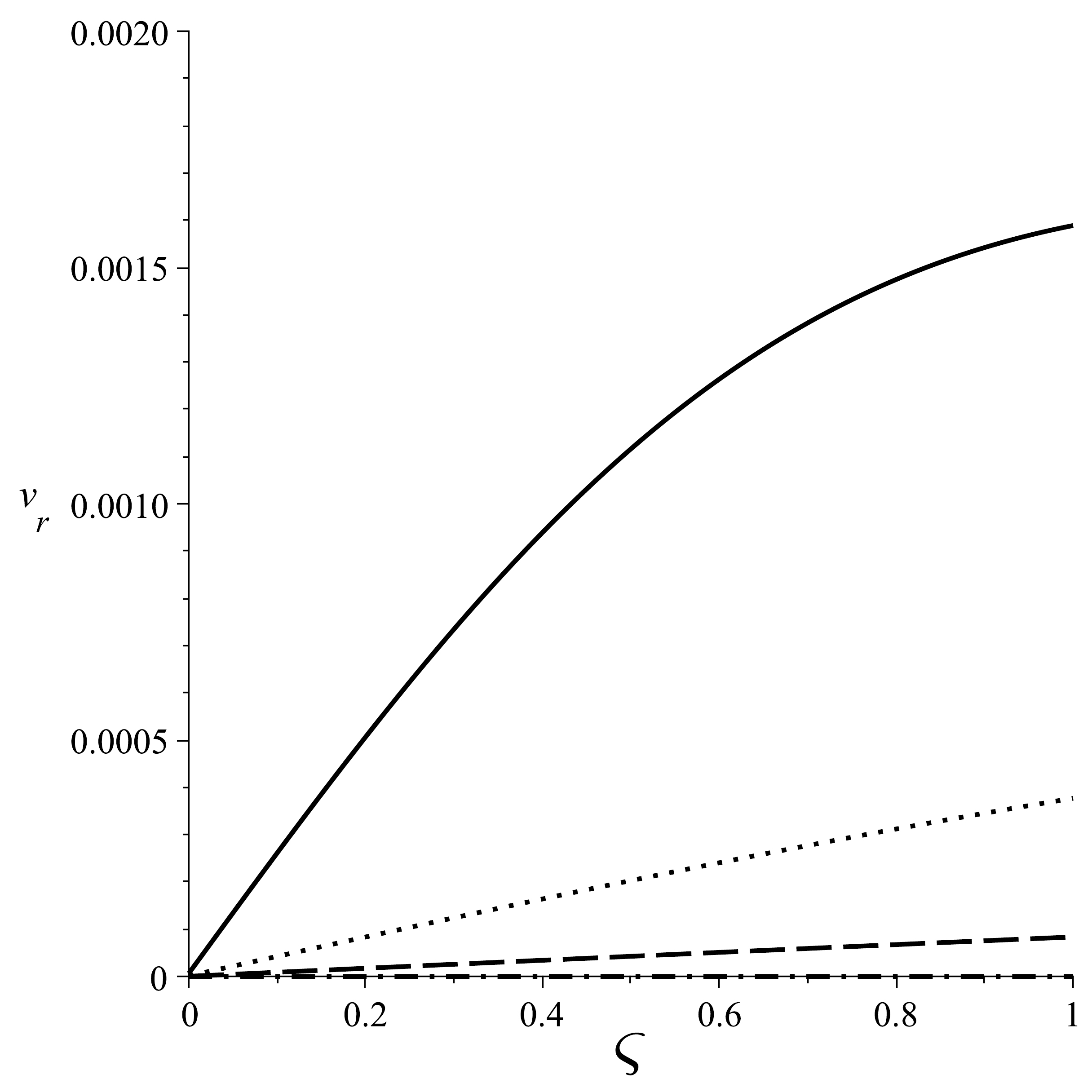}
        \centering
        \caption{}
        \label{fig:vrr}
    \end{subfigure}
    ~ %add desired spacing between images, e. g. ~, \quad, \qquad, \hfill etc. 
    \begin{subfigure}[b]{0.485\textwidth}
        \includegraphics[width=\textwidth]{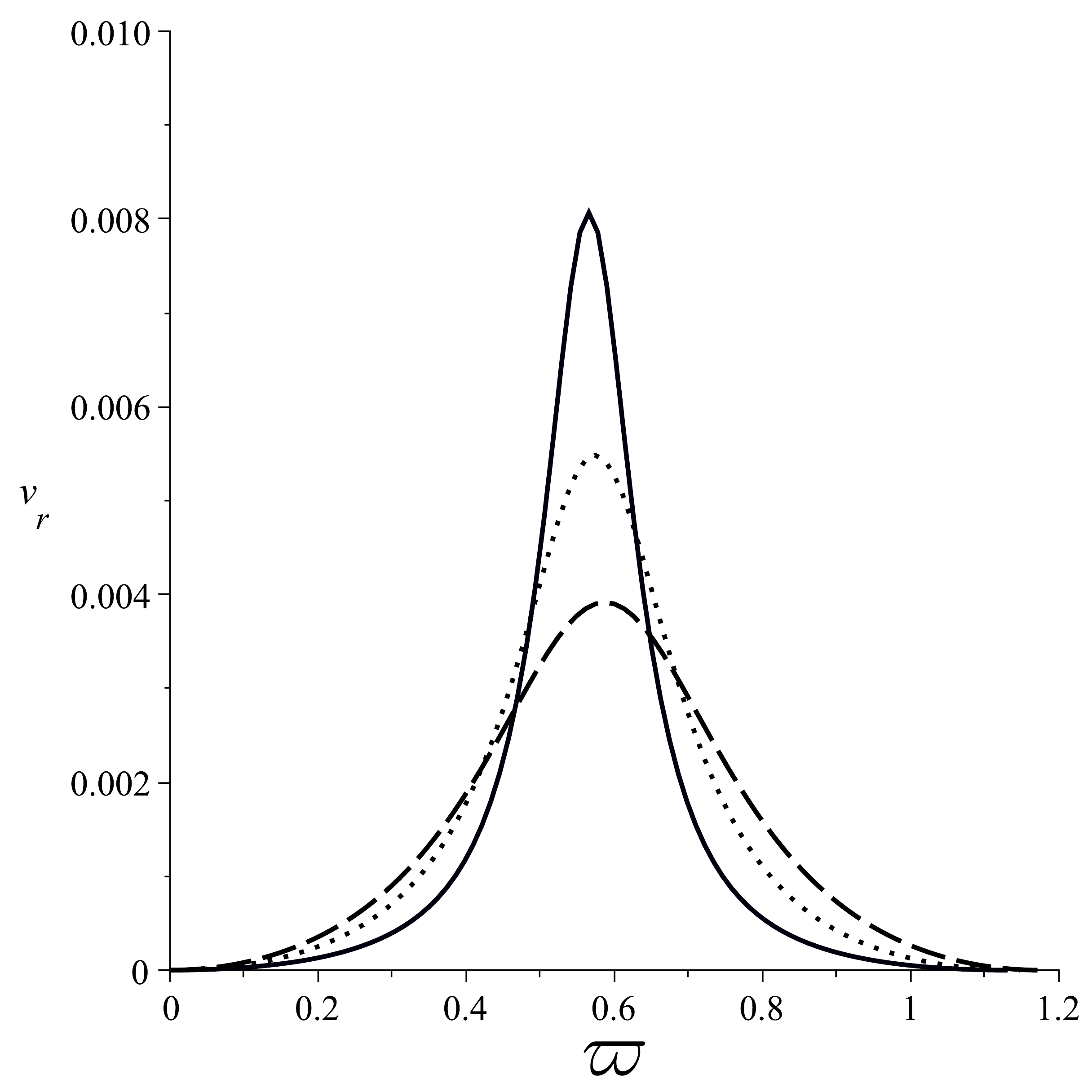}
        \centering
        \caption{}
        \label{fig:vrz}
    \end{subfigure}
    \caption{Normalized radial peculiar velocity $v_r$ at present time, where the energy flux is interpreted as peculiar velocities $q_a=\rho v_a$ for dust matched with a Szekeres--II section with energy flux. Panel (a) displays $v_r$ vs. $\varsigma=cSr/\bar H_0$  for various values of $w$. Panel (b) displays $v_r$ vs. $\varpi=S\int X dw$ for various values of $r$.}\label{fig:vrSz}
\end{figure}

We display in figures \ref{fig:OmSz} and \ref{fig:vrSz} the density $\hat\Omega$ and the radial peculiar velocities at present time $S(\tau_0)=1$ as functions of the proper distances, $\varsigma=cSr/\bar H_0$ for the $r$ direction and $\varpi=cS\int X dw/\bar H_0$ for the $w$ direction. The matchings interfaces with Einstein de Sitter sections are marked by $w=0$ and $w= 2\pi/\nu_0$, where the free functions are as stated in the previous paragraph and $\nu_0=\nu$. In figures \ref{fig:Omr0} and \ref{fig:vrr} we selected four values of the $w$ coordinate: $w=\pi/2\nu_0, \pi/\nu_0,3\pi/2\nu_0, 2\pi/\nu_0$, respectively depicted by solid, dotted, dashed and dash dot curves. Likewise, figures \ref{fig:Omz0} and \ref{fig:vrz} we selected three values of $r$: $\varsigma=1/3, 2/3, 1$, respectively represented by  solid, dotted and dashed curves. It is straightforward to verify that the matching conditions (\ref{1junc})--(\ref{3junc}) hold. The velocities are plotted as fractions of the speed of light, Notice that the orders of magnitude are as expected, $|v_r|<2400 \, \textrm{km}/\textrm{s}$. 

\section{Exact perturbations on an FLRW background}\label{perturbations}

The examples we have provided of multiple matchings between Szekeres--II and FLRW regions describe multiple localized inhomogeneities in a homogeneous and isotropic background defined by these matchings (not by a sequence of models converging to FLRW geometry). As such, these configurations can be interpreted as exact perturbations on these backgrounds. We can phrase this notion in rigorous covariant terms by looking at kinematic and curvature quantities. For the examples we examined in the previous section the electric and magnetic Weyl tensores are given by
\begin{eqnarray}
E^a_b = {\cal E}\,\xi^a_b,\quad {\cal E} = \frac16\xi_{ab}E^{ab}=-\frac{\kappa}{6}\left(\rho-\bar\rho\right),\qquad
H_{ab}=\frac{\kappa}{2}\eta_{abc}q^c,
\end{eqnarray}
where $\xi^a_b$ is the same traceless tensor eigenframe as the shear tensor in (\ref{eq:eqThSig}) and $\kappa\bar\rho=3\dot S^2/S^2=3\dot a^2/a^2$ is the FLRW density. As a consequence (and using (\ref{eq:eqThSig})), we can express the dimensionless density,  Hubble scalar and peculiar velocities of the Szekeres--II regions as exact perturbations over the parameters of the FLRW regions given in terms of the eigenvalues of the electric Weyl and shear tensor and in terms of the magnetic Weyl tensor
\begin{eqnarray}
\fl\hat\Omega = \bar\Omega +\delta^\Omega,\quad \delta^\Omega=-\frac{\xi_{ab}E^{ab}}{3\bar H_0^2},\qquad \hat {\cal H} = \bar {\cal H} + \delta^{\cal H},\quad \delta^{\cal H}=-\frac{\xi_{ab}\sigma^{ab}}{6\bar H_0},\label{exactp1}\\
\fl \Rightarrow\qquad  E_{ab}  = -\frac12\delta^\Omega\,\bar H_0^2\,\xi_{ab},\qquad \sigma_{ab}=-\delta^{{\cal H}}\,\bar H_0\,\xi_{ab},\label{exactp2}\\
\fl H_{ab}=\frac32\hat\Omega\,\bar H_0^2\,\eta_{abc}v^c=\frac32\left(\bar\Omega+\delta_\Omega\right)\,\bar H_0^2\,\eta_{abc}v^c,\qquad \hat\Omega^a_q\equiv \frac{\kappa\,q^a}{3\bar H_0^2}=\hat\Omega\,v^a=\left(\bar\Omega+\delta_\Omega\right)\,v^a,\label{exactp3} 
\end{eqnarray}
where $\xi_{ab}$ is defined in (\ref{eq:eqThSig}), the FLRW parameters $\bar\Omega$ and $\bar {\cal H}$ have been defined and computed:
\begin{eqnarray}\fl \bar\Omega =\frac{\Omega_0^m}{a^3}+\Omega_0^\Lambda\,\,\,(\Lambda\hbox{CDM}),\quad \Omega_0^\Lambda\,\,\,(\hbox{de Sitter}).\quad \frac{4}{9\tau^2}\,\,\,(\hbox{Einstein de Sitter}), \nonumber\end{eqnarray}
with $\bar {\cal H} = \sqrt{\Omega}$ in all cases. Although the density and Hubble scalar $\hat\Omega,\,\hat{\cal H}$ of the Szekeres--II regions have been expressed already in the form of background FLRW values plus extra quantities that depend on all coordinates in (\ref{rhoLCDM})--(\ref{HLCDM}), (\ref{ecrhodS}) and (\ref{eqhatOM})--(\ref{eqhatH}), equations (\ref{exactp1})--(\ref{exactp2}) relate these quantities to the shear and electric Weyl tensors. 

Although the FLRW background is spatially flat, the Szekeres--II regions are not spatially flat: the Ricci scalar of the hypersurfaces orthogonal to $u^a$ is in general not zero.  Hence, the spatial curvature becomes also an exact perturbation that vanishes at the matching hypersurface. The spatial curvature perturbation expressed in terms of the variables in (\ref{exactp1}) is 
\begin{equation} \delta^K\equiv \frac{{}^{(3)}{\cal R}}{6\bar H_0^2}= \delta^\Omega-2\bar{\cal H}\delta^{\cal H} = \frac{\alpha_3 + Q\beta_3}{3\bar H_0^2\,S^2\,X},\label{curvpert}\end{equation}
where the relation between $\delta^K$ and the other perturbations follows from the Hamiltonian constraint ${}^{(3)}{\cal R}=2\kappa\rho-(2/3)\Theta+\sigma_{ab}\sigma^{ab}$.   

Regarding the peculiar velocity vector $v^a$, it is the only approximate ({\it i.e.} linear) perturbation, as we are assuming $v^a\ll 1$ and neglecting second order terms to identify $q^a=\rho v^a$. Notice that these velocities are absent in the two examples of pure comoving dust for which the Szekeres--II solution is Petrov type D and thus $H_{ab}=0$. The fulfillment of the matchings conditions (\ref{21junc})--(\ref{3junc}) guarantees that all the exact perturbations (as well as $v^a$) smoothly vanish at the matching interface with the FLRW sections.

The role of $\delta_\Omega,\,\delta_{\cal H}$ as covariant exact perturbations (and $v^a$ as a covariant linear perturbation) on an FLRW background characterized by $\bar\Omega=\bar{\cal H}^2$ (with ${\cal H}=2/(3\tau)$) can be expressed rigorously through  their evolution equations obtained from (\ref{mombal1})-(\ref{mombal2}) and (\ref{Raych}). Since $\delta^\Omega,\,\delta^{{\cal H}}$ and $v^a$ are directly related to the electric Weyl, shear and magnetic Weyl tensors, we have (up to first order in $v^a$)   
\begin{eqnarray}
\fl\delta^{\cal H}_{,\tau}=-2\bar{\cal H}\,\delta^{\cal H}-3[\delta^{\cal H}]^2-\frac12\delta^\Omega,\label{perturb1}\\
\fl\delta^\Omega_{,\tau}= -3(\bar{\cal H}+\delta^{\cal H})\,\delta^\Omega-\bar\Omega\delta^{\cal H}+\bar\Omega\left(\frac{v^{\tilde x}\,X_{,\tilde x}}{X}+\frac{v^{\tilde y}\,X_{,\tilde y}}{X}\right)-(\bar\Omega+\delta^\Omega)(v^{\tilde x}_{,\tilde x}+v^{\tilde y}_{,\tilde y}),\label{perturb2}\\
\fl v^{\tilde x}_{,\tau}=v^{\tilde x}\,\left[v^{\tilde x}_{,\tilde x}+v^{\tilde y}_{,\tilde y}-2\bar{\cal H}\right],\qquad v^{\tilde y}_{,\tau}=v^{\tilde y}\,\left[v^{\tilde x}_{,\tilde x}+v^{\tilde y}_{,\tilde y}-2\bar{\cal H}\right], \label{perturb3}
\end{eqnarray}
where $\tilde x=\bar H_0\,x$ and $\tilde y=\bar H_0\,y$ and we used the constraint $\tilde\nabla_b E^b_a-\kappa/3[h_a^b\rho_{,b}+q_a+(3/2)\sigma_{ab}q^b]=0$ at first order in $q_a=\rho v_a$ to arrive to equations (\ref{perturb2})--(\ref{perturb3}).  
The comoving dust case in the two examples examined before ($\Lambda$CDM and de Sitter backgrounds) follows from (\ref{perturb1})--(\ref{perturb3}) by setting $v^a=0$. For this case it is interesting to combine (\ref{perturb1})--(\ref{perturb2}) into a second order non--linear equation for the density perturbation 
\begin{eqnarray}
\fl\delta^\Omega_{,\tau\tau}-3(\bar{\cal H}+\delta^{\cal H})\delta^\Omega_{,\tau}+\frac32[\delta^\Omega]^2+\frac32\left[\bar\Omega+4\left(\bar{\cal H}+\frac32\delta^{\cal H}\right)\delta^{\cal H}\right]\delta^\Omega+6\bar\Omega\left(\bar{\cal H}+\frac32\delta^{\cal H}\right)\delta^{\cal H}=0,\nonumber\\
\end{eqnarray}
that becomes in the linear limit
\begin{equation}\delta^\Omega_{,\tau\tau}-3\bar{\cal H}\delta^\Omega_{,\tau}+\frac92\bar\Omega\delta^\Omega -3\bar\Omega\delta^K=0,\end{equation}
where we used (\ref{curvpert}). This linear equation coincides with dust linear perturbations in the comoving gauge save for the last term proportional to the curvature perturbation $\delta^{\cal K}$. The fact that we do not recover the usual linear dust perturbation follows from the fact that the FLRW background is defined by a matching instead of by a sequence of models converging to an FLRW geometry. From the functional form of $\delta^{\cal K}$ in (\ref{curvpert}) for the comoving case $Q=0$, we have $\delta^{\cal K}=0$ only if we select models such that $\alpha_3(w)=0$ (the coefficient of the quadratic terms $x^2+y^2$ in the Szekeres-II dipole in (\ref{noPIyesQ2})). As shown in \cite{DelBuch} (see their Appendix C) the vanishing of this coefficient is a necessary condition for an FLRW limit in the parameter space.

\section{Discussion and conclusions}\label{sec:disc}

We have presented a novel approach to Szekeres--II models that have been regarded as unsuitable to describe the evolution of cosmological inhomogeneities in a homogeneous background. We have devised ``pancake'' configurations constructed by smoothly matching quasi-flat Szekeres--II regions with any cosmology compatible with the Robertson--Walker metric: FLRW models, but also de Sitter, anti de Sitter and Minkowski spacetimes (section \ref{sec:junc}). As shown in section \ref{sec:SzekII}, Szekeres--II models are (in general) Petrov type I and thus are compatible with a wide variety of sources, including mixtures of fluids and scalar fields and with sources that require vector and tensor modes, such as magnetic fields and gravitational waves.   

The resulting ``pancake'' constructions described above allow for a description of an arbitrary countable number of localized inhomogeneities evolving together, embedded in a homogeneous and isotropic cosmology to which they are smoothly matched. We provided in section \ref{sec:example}  three simple toy examples to illustrate the ``pancake'' configurations: Szekeres--II dust regions in $\Lambda$CDM and de Sitter backgrounds and regions of non--comoving dust in an Einstein de Sitter background. As we show in section \ref{perturbations}, the eigenvalues of the electric Weyl and shear tensors respectively constitute the inhomogeneous part of the density and Hubble scalar, while the peculiar velocities (in the example considering them) are directly related to the magnetic Weyl tensor. Since these inhomogeneous parts vanish at the FLRW or de Sitter matching, the Szekeres--II inhomogeneities can be rigorously considered as exact covariant perturbations on the background defined by these matchings.   

It is worthwhile relating the ``pancake'' constructions we have devised and discussed here with the periodic lattice models presented by Delgado and Buchert in \cite{DelBuch}, also based on similar matchings but only involving comoving Szekeres-II dust regions and dust FLRW spacetimes (they only considered the Einstein de Sitter model as they assumed $\Lambda=0$). These authors considered Szekeres-II dust models to probe a formalism they developed to generate a fully general relativistic generalization of the Newtonian Zeldovich approximation. As they show, the smooth matching with FLRW regions is a necessary and sufficient condition for the existence of ``homogeneity domains'' for which the kinematic backreaction (in Buchert's averaging formalisms) vanishes identically. Since the ``pancake'' constructions we have introduced in this article are more general that just comoving dust with $\Lambda=0$, allowing for a wide variety of sources and FLRW regions and not restricted to have a periodic lattice distribution, it is certainly worthwhile to further probing the results of \cite{DelBuch}. 

Hoping to motivate further exploration of these configurations we are currently working on less idealized configurations involving realistic peculiar velocities and a $\Lambda$CDM background \cite{Noncom}. While these models are inherently inhomogeneous and anisotropic, given their richness of free parameters the challenge is to accommodate their inhomogeneity and anisotropy to fit observational constraints. Finally, the possibility of describing the dynamics of mixtures of fluids with scalar fields, magnetic fields and gravitational waves embedded in de Sitter or anti de Sitter spacetimes, as thick 4--dimensional branes, leads to potential applications of Szekeres--II models to early universe modeling, including inflationary scenarios,  reheating and semi--classical quantum field theory. These possible applications are worth looking at in future work

\appendix

\section{Compatibility conditions}\label{sec:Compa}

Given a reference $4$-velocity the energy-momentum tensor of an \emph{imperfect fluid} is given by 
\begin{equation}\label{eq:Tab}
T^{ab} = \rho u^a u^b + p h^{ab}+ \pi^{ab}+ 2 q^{(a} u^{b)} 
\end{equation}
where $h^{ab}=u^a u^b+g^{ab}$ is the projection operator and the following quantities 
$$ \rho=u_a u_bT^{ab},\quad p =\frac13 h_{ab} T^{ab},\quad \pi^{ab}=T^{\langle ab\rangle}=\left[h^{(a}_ch^{b)}_d-\frac13 h^{ab}h_{cd}\right]\,T^{cd} ,\quad q_a = -u^b T_{ab}$$
are the mass--energy density ($\rho$), the isotropic pressure ($p$), anisotropic pressure ($\pi^{ab}$) and the energy flux relative to the $4$-velocity ($q_a$). \\

We consider two general observers in spacetime, which have different 4-velocities $u^a$ and $\hat{u}^a$. Choosing $u^a$ as the reference $4$-velocity, $\hat{u}^a$ is related to this reference $4$-velocity as $\hat{u}^a=\gamma(u^a+v^a).$ The energy-momentum tensor of the dust source in the frame of the observer is
\begin{eqnarray}
\tensor{\hat{T}}{_a_b}=\rho \hat{u}_{a} \hat{u}_b =\rho\gamma^2 \left( u_a u_b +2 u_{(a}v_{b)}+v_a v_b\right).
\end{eqnarray}
Decomposing the energy-momentum tensor can be decomposed in its parts parallel,
mixed and orthogonal to $u_a$
\begin{equation}
T_{ab}=\rho u_a u_b + p h_{ab} + 2u_{(a}q_{b)}+\Pi_{ab}$$	
\end{equation}
where we now consider $p=0$, $\Pi_{ab}=0$ and search under what conditions, in the limit $v_av^a\to 0$, $T_{ab}=\tensor{\hat{T}}{_a_b}$. To order zero, \cite{Ellisgen}, $\lim_{v_av^a} \tensor{\hat{T}}{_a_b} = \rho\left( u_a u_b +2 u_{(a}v_{b)}\right)$, therefore the first condition would be $\rho u_{(a}v_{b)}=u_{(a}q_{b)}$. 

Even though we take $v_av^a\ll 1$, this does not imply the derivatives are small, so we must search conditions to first order. As $\tensor{\hat{T}}{^a^b_;_b}=0$ and $\tensor{T}{^a^b_;_b}=0$, the difference 
\begin{equation}
\tensor{\hat{T}}{^a^b_;_b}-\tensor{T}{^a^b_;_b}=0 \label{eq:derdiff}
\end{equation}
will yield the second condition when we obtain an identity $0=0$ in the limit considered. As
\begin{eqnarray}
\tensor{q}{^a_;_b} &=& \rho_{;b}\gamma^{2}v^a+2\rho\gamma \gamma_{;b}v^a + \rho\gamma^{2}\tensor{v}{^a_;_b},
\label{qder}\\
\rho_{;b} &=&\rho_{;b}\gamma^2+ 2\rho\gamma \gamma_{;b},
\label{rhoder}
\end{eqnarray}

It is straightforward to verify that zero order relations together with (\ref{eq:derdiff})-(\ref{rhoder}) yield 
\begin{equation}
\rho\gamma^2(v^a v^b)_{;b}+(\rho_{;b}\gamma^{2}+2\rho\gamma\gamma_{;b})v^a v^b=0\label{eq:subs2}.
\end{equation}
As previously stated, we neglect quadratic terms on $v^a$. Therefore our second condition will be:
\begin{equation}
(v^a v^b)_{;b}=0\label{eq:2order}
\end{equation}

This implies $v^a v^b$ is constant, which we take as $v^a v_a <<1$ for consistency with our initial hypothesis $v^a v_a <<1$. Therefore our conditions for compatibility are 
\begin{eqnarray}
\rho u_{(a}v_{b)}&=&u_{(a}q_{b)},\\
(v^a v^b)_{;b}&=&0.
\end{eqnarray}
With this considerations we calculated the energy conservation equation $u_a\tensor{T}{^a^b_;_b}$ and obtained a function proportional to $(H_0/c)^3$ which justifies our approximation. 

\section*{Acknowledgements}
SN acknowledges financial support from SEP–-CONACYT postgraduate grants program and RAS acknowledges support from PAPIIT--DGAPA RR107015.

\section*{References}
\bibliographystyle{iopart-num}
\bibliography{MatchingCQG}

\providecommand{\newblock}{}
\begin{thebibliography}{10}
\expandafter\ifx\csname url\endcsname\relax
  \def\url#1{{\tt #1}}\fi
\expandafter\ifx\csname urlprefix\endcsname\relax\def\urlprefix{URL }\fi
\providecommand{\eprint}[2][]{\url{#2}}

\bibitem{Plebanski}
Plebanski J and Krasinski A 2006 {\em An introduction to general relativity and
  cosmology\/} (Cambridge University Press)

\bibitem{Krasinski}
Krasi{\'n}ski A 2006 {\em Inhomogeneous cosmological models\/} (Cambridge
  University Press)

\bibitem{Sussbol}
Bolejko K and Sussman R~A 2011 {\em Physics Letters B\/} {\bf 697} 265--270

\bibitem{SussDel}
Sussman R~A and Gaspar I~D 2015 {\em Physical Review D\/} {\bf 92} 083533

\bibitem{Kasai1}
Kasai M 1992 {\em Physical review letters\/} {\bf 69} 2330

\bibitem{Ishak1}
Ishak M and Peel A 2012 {\em Physical Review D\/} {\bf 85} 083502

\bibitem{Ishak2}
Peel A, Ishak M and Troxel M 2012 {\em Physical Review D\/} {\bf 86} 123508

\bibitem{SusQue}
Quevedo H and Sussman R~A 1995 {\em Classical and Quantum Gravity\/} {\bf 12}
  859

\bibitem{Coll}
Coll B, Ferrando J~J and Saez J~A 2020 {\em Classical and Quantum Gravity\/}

\bibitem{DelBuch}
Gaspar I~D and Buchert T 2020 Lagrangian theory of structure formation in
  relativistic cosmology. vi. comparison with szekeres exact solutions
  (\textit{Preprint} \eprint{arXiv:2009.06339})

\bibitem{Ellis}
Ellis G~F, Maartens R and MacCallum M~A 2012 {\em Relativistic cosmology\/}
  (Cambridge University Press)

\bibitem{Larena1}
Bruneton J~P and Larena J 2012 {\em Classical and Quantum Gravity\/} {\bf 29}
  155001

\bibitem{Clifton}
Clifton T, Ferreira P~G and O’Donnell K 2012 {\em Physical Review D\/} {\bf
  85} 023502

\bibitem{Cliftonback}
Clifton T 2015 Back-reaction in relativistic cosmology {\em THE THIRTEENTH
  MARCEL GROSSMANN MEETING: On Recent Developments in Theoretical and
  Experimental General Relativity, Astrophysics and Relativistic Field
  Theories\/} (World Scientific) pp 2553--2565

\bibitem{Israel}
Israel W 1966 {\em Il Nuovo Cimento B (1965-1970)\/} {\bf 44} 1--14

\bibitem{Mars}
Mars M and Senovilla J~M 1993 {\em Classical and Quantum Gravity\/} {\bf 10}
  1865

\bibitem{Goode}
Goode S~W 1986 {\em Classical and Quantum Gravity\/} {\bf 3} 1247

\bibitem{Bruni}
Meures N and Bruni M 2011 {\em Physical Review D\/} {\bf 83} 123519

\bibitem{Ellisgen}
Maartens R, Gebbie T and Ellis G~F 1999 {\em Physical Review D\/} {\bf 59}
  083506

\bibitem{Noncom}
N\'ajera S and Sussman R~A 2020 In preparation Non-comoving cold dark matter in
  a {$\Lambda$CDM} background

\end{thebibliography}

\end{document}